# Uncertainty and sensitivity analysis of hair growth duration in human scalp follicles under normal and alopecic conditions

Atanaska Dobreva[1]*, Damon Comer[1], N.G. Cogan[2], Ralf Paus[3]
[1] Augusta University, Augusta, GA, USA
[2] Florida State University, Tallahassee, FL, USA
[3] University of Miami Miller School of Medicine, Miami, FL, USA

## Abstract

Hair follicles, the organs which produce hair, go through a constant cycle composed of phases of growth, regression and rest, as matrix keratinocytes (MKs), the cells responsible for hair fiber synthesis proliferate for several years, and then undergo spontaneous apoptosis. Damage to MKs and perturbations in their normal dynamics result in a shortened growth phase of the hair cycle, leading to hair loss. The most common factors causing such disruption are hormonal imbalance and attacks by the immune system. Androgenetic alopecia (AGA) is a form of hair loss caused by high sensitivity to androgens, and alopecia areata (AA) is a condition where hair loss is caused by an autoimmune reaction against MKs. In this study, we inform a mathematical model for the human hair cycle with experimental data for the lengths of hair cycle phases available from male control subjects and subjects with AGA. We also, connect a mathematical model for AA with estimates for the duration of hair cycle phases obtained from the literature. Subsequently, with each model we perform parameter screening, uncertainty quantification and global sensitivity analysis and compare the results within and between the control and AGA subject groups as well as among AA, control and AGA conditions. The findings reveal that in AGA subjects there is greater uncertainty associated with the duration of hair growth than in control subjects and that, compared to control and AGA conditions, in AA it is more certain that longer hair growth phase could not be expected. The comparison of results also indicates that in AA lower proliferation of MKs and weaker communication of the dermal papilla with MKs via signaling molecules could be expected than in normal and AGA conditions, and in AA stronger inhibition of MK proliferation by regulatory molecules could be expected than in AGA. Finally, the global sensitivity analysis highlights the process of MK apoptosis as highly impactful for the length of hair growth only in the AA case, but not for control and AGA conditions.

## Introduction

Hair follicles (HFs), one of the defining features of mammals, undergo a life-long, spontaneous organ remodeling process, the hair cycle (HC). The hair cycle is most prominently exhibited in the continuous cyclical apoptosis of the epithelial cells directly responsible for hair shaft production, the matrix keratinocytes (MKs), during HF involution (catagen) and their differentiation and proliferation from resident epithelial progenitor cell pools during anagen development [1, 2]. The autonomous, intrafollicular oscillator system that drives this rhythmic process of HF regression and growth, the HC clock (HCC), is a complex biological clock mechanism that unfolds over a very long period of time, i.e. on the order of years in human scalp HFs [3, 4]. While the molecular nature of the human HCC remains to be fully elucidated, HC-dependent changes in intrafollicular peripheral clock gene activity within MKs likely play a key role [5], even though their hierarchy within the HCC and coordination with other HCC-regulatory elements recognized in murine HC biology such as MPZL3 [6] as well as the central generator of rhythmicity within the HCC and its precise anatomical location within the HF are still quite unclear. To shed light on the elusive HCC and to interrogate the dynamics of HF cycling, several attempts have been made to use mathematical modeling [7, 8, 9, 10, 11, 12, 13]. These previous studies have focused on HF cycling in mice, not humans.

---

* Corresponding author: adobreva@augusta.edu

Nagorcka (1983) [8] and Nagorcka and Mooney (1982) [7] proposed a reaction-diffusion system to describe control of cellular differentiation in the HF bulb by the interactions of three morphogens, but the model does not explain the cycling of HFs. Golichenkova and Doronin (2008) created a model for hair volume, where the parameters were informed using data on rate of hair elongation and information about changes in the geometry of the hair fiber during elongation [9]. However, the model cannot elucidate mechanisms underlying the cyclical nature of HF regeneration. Plikus et al. (2011) examined hair growth and cycling in mice [10], and Murray et al. (2012) modeled synchronized follicle growth in mice as an excitable medium, showing qualitative agreement to experimental observations [11]. Tasseff et al. (2014) analyzed large scale transcriptome-wide expression data sets from mice using a dynamic coupled oscillator model and identified two clusters of genes that exhibited synchronized, out of phase expression profiles, concluding that the synchronization likely involves a balance of negative and positive feedback [12]. Wang et al. (2017) developed a modeling framework for murine HFs that associates the dynamics of activator and inhibitor molecules with cyclic growth in a single HF and captures coupling between multiple HFs using diffusive signals [13]. However, growth is described with an abstract HF length variable devoid of information about the dynamics of MKs.

However, there is a critical difference between the features of hair follicle cycling in mice compared to humans in that human hair follicles cycle independently (mosaic HC pattern) [14]. Moreover, in striking contrast to mice, human scalp HFs stay for several years in anagen and then run through a very short catagen and telogen transformation before they return into anagen, in an almost bistable manner [4]. Therefore, HC modeling must be species-specific, concepts and models that reflect murine HF cycling, may not be applicable to the human system.

Al-Nuaimi et al. (2012) proposed a kinetic mathematical model for cyclical hair follicle growth patterns in humans with negative feedback between a static population of dermal papilla cells and a dynamic population of MKs [14]. Apart from the model of Al Nuaimi et al. (2012), there is only one other model for the human hair cycle by Halloy et al. (2000) [15], (2002) [16] which is of cellular automaton type. However, cellular automaton models for oscillator systems are not driven by the mechanisms and cellular interactions underlying each oscillation, so this modeling approach does not allow for examining the direct connection between cellular properties and globally observed behavior [17, 18]. The advantage of a mechanistic model is that it enables the gathering of insight into the direct relationship between cellular properties and global behavior expressed in the hair cycle. The hair cycle model from Al-Nuaimi et al. (2012) is the only mechanistic model for the human hair cycle, and it connects dynamics at the molecular and cellular level, showing oscillations in the number of MKs [14].

Another challenge that any good model of human HF cycling dynamics must take into account is that MKs are highly sensitive to damage by various factors, and perturbations in their normal dynamics lead to undesired alteration of hair follicle cycling in the form of a shortened growth phase (anagen), which results in hair loss [3, 19]. Alopecia areata (AA) is one of the most common autoimmune diseases, where damage is caused by inflammatory mediators, and we have performed studies interconnecting dermatology and mathematics and published mathematical models that capture key features of the autoimmune response and subsequent hair loss symptoms in AA [20, 21, 22, 23].

With an ordinary differential equation (ODE) model, we first captured interactions between follicles in phase of growth and immune components that experimental studies classify as drivers of AA development: $CD8^+$ T-cells, $CD4^+$ T-cells, the pro-inflammatory cytokine interferon-gamma (IFN-$\gamma$), and immune privilege guardians (IPG) that have immunosuppressive activity. Linear stability and bifurcation analysis revealed that that changes in IFN-$\gamma$ cause a transcritical bifurcation, where as the amount of IFN-$\gamma$ increases past a critical value, the state of health loses stability and the state of disease, marked by very high levels of the immune cells, becomes stable [20]. In our next study, we linked this model for immune components with the hair cycle equations from Al-Nuaimi et al. (2012). The composite system demonstrates hair growth disruption in response to the autoimmune reaction in AA, where the disease profile is manifested through a very short growth phase as the immune cell populations expand in response to elevated IFN-$\gamma$ [21].

We also developed a partial differential equation (PDE) model, which captures the distinct patterns of hair loss in AA. The system describes the temporal and one-dimensional spatial dynamics of the key immune components. The immunocyte dynamics include activation by IFN-$\gamma$, proliferation (both being suppressed by IPG), and death. The cells also exhibit random motion as well as motion directed up the concentration gradient in IFN-$\gamma$. The numerical simulations of the PDE model illustrate the development of a pattern characteristic of AA with spatial and temporal scale consistent with those reported in experimental findings [22]. Subsequently, we used a Bayesian framework based on data assimilation to examine how parameter sensitivity as well as frequency and noise in observational data collection affect the estimation of parameters in the PDE model. We found that estimates of sensitive parameters converged faster than estimates for insensitive parameters, and our results showed that sensitive parameters are affected by noise to a much greater extent, and their estimates become less accurate as the noise level increases [23].

Another common form of hair loss is androgenetic alopecia (AGA), also known as pattern baldness, where hair loss occurs due to other factors, such as hormonal imbalance. AGA is characterized by the presence of areas on the scalp where hair follicles have high sensitivity to androgens, and the sizes and locations of these areas depend on genetic factors. Alteration of the hair cycle occurs with the length of anagen progressively shortening, and this together with hair fiber miniaturization leads to the formation of scalp regions with decreased or no hair coverage [24]. Courtois et al. (1995) is the only experimental study, which has followed hair growth in human subjects over a period of at least 8 years in order to measure the lengths of the anagen and telogen phases of the hair cycle [25]. These observations were performed in male control subjects and subjects with AGA.

As mathematical models have not yet been used to perform model-based analysis of experimental data from human subjects for the duration of the hair cycle phases, in this study we connect the Al-Nuaimi et al. (2012) model with the data supplied by the Courtois et al. (1995) study for the lengths of anagen and telogen in male subjects with and without AGA. Through modulation of the parameter in the model that governs MK apoptosis, we are able to achieve alignment between the data and the model simulation prediction for each subject. We then combine uncertainty quantification and sensitivity analysis, which equips us with a unique way to compare within and between the subject groups in order to investigate important aspects of the heterogeneity in the lengths of anagen in normal conditions versus AGA conditions. In addition, we perform the steps of model simulation alignment, uncertainty quantification and sensitivity analysis to explore the case of AA. For this purpose, we use our previously proposed model in Dobreva et at. (2018) for immune system dynamics and the development of AA in cycling hair follicles [21], and we compare the results with normal and AGA conditions. For the AA case, estimates for the lengths of anagen and telogen are obtained through deduction from information available in the literature.

## Methods

### Mathematical models

#### Model for the hair cycle

Al-Nuaimi et al. (2012) proposed a kentic mathematical model for the human hair cycle motivated by follicular mechanisms and interactions underlying the cycling of a single follicle. The model comprising Equations (1)-(5) below, captures interactions between two compartments matrix keratinocytes (MKs) and dermal papilla (DP), and it predicts the lengths of time spent in each phase of the hair cycle. Equation (1) describes dynamics at the cellular level, namely the population dynamics of MKs over time ($\xi$). Equations (2)-(5) describe dynamics at the molecular level, specifically changes over time in the amounts of signaling and regulatory molecules in the MK population compartment and the DP compartment. The model variables $\eta_1$ and $\eta_2$ are the amount of signaling molecules in the MK population compartment and the DP compartment, respectively, and the model variables $z_1$ and $z_2$ are the amounts of regulatory molecules in the MK population compartment and the DP compartment, respectively. The model output is a relaxation oscillation curve, where the long upper state is interpreted as anagen, the lower short state as telogen and the fast transition as catagen [14].

$$\frac{d\xi}{dt} = \frac{p_1\xi}{(p_2+\xi)(p_3+C_{prol}z_1)} - \frac{p_4\xi}{p_5^k+\xi^k} + \alpha - \beta\xi \tag{1}$$

$$\frac{d\eta_1}{dt} = c_1\xi + D_\eta(\eta_2 - d_1\eta_1) \tag{2}$$

$$\frac{d\eta_2}{dt} = D_\eta(\eta_1 - d_2\eta_2) \tag{3}$$

$$\frac{dz_1}{dt} = D_z(z_2 - d_3z_1) \tag{4}$$

$$\frac{dz_2}{dt} = c_2\eta_2 + D_z(z_1 - d_4z_2) \tag{5}$$

## Model for AA

In our previous study Dobreva et at. (2018) [21], we proposed a model for AA with hair cycling by combining in a composite system the model of ordinary differential equations we had developed for the autoimmune response in AA [20] with the Al-Nuaimi et al. (2012) hair cycle model [14]. The immune sub-system captures the temporal dynamics of immune cells considered principal drivers of AA development: CD8+ T-cells ($x$) and CD4+ T-cells ($y$). Through the model parameters, the equations also reflect the impact of the pro-inflammatory cytokine interferon-gamma (IFN-$\gamma$) as well as the impact of immune privilege guardians, which are locally produced substances that suppress immune activities [20]. The immune sub-system is connected with the hair cycle model, as shown below, through a function $f(x, y)$, which feeds the output of the immune sub-system into the apoptosis term of the equation for MKs [21].

*Sub-system for the dynamics of immune cells*

$$\frac{dx}{dt} = \frac{v(x+y)}{(1+s)(a+y+x)} + \frac{lxy}{(1+s)} - x - x^2 \tag{6}$$

$$\frac{dy}{dt} = \frac{b(x+y)}{(1+s)} + \frac{wy}{(1+s)} - y - y^2 \tag{7}$$

*Equations for the hair cycle in the presence of AA*

$$f(x,y) = \lambda(x+y) + 1 \tag{8}$$

$$\frac{d\xi}{dt} = \frac{p_1\xi}{(p_2+\xi)(p_3+C_{prol}z_1)} - \frac{\boxed{f(x,y)}\,p_4\xi}{p_5^k+\xi^k} + \alpha - \beta\xi \tag{9}$$

$$\frac{d\eta_1}{dt} = c_1\xi + D_\eta(\eta_2 - d_1\eta_1) \tag{10}$$

$$\frac{d\eta_2}{dt} = D_\eta(\eta_1 - d_2\eta_2) \tag{11}$$

$$\frac{dz_1}{dt} = D_z(z_2 - d_3z_1) \tag{12}$$

$$\frac{dz_2}{dt} = c_2\eta_2 + D_z(z_1 - d_4z_2) \tag{13}$$

# Experimental Data

## AGA

In this study we connect Al-Nuaimi et al. (2012) model for the hair cycle with experimetal data on the lengths of anagen and telogen that Courtois et al. (1995) collected from male subjects without and male subjects with androgenetic alopecia. Courtois et al. (1995) used the non-invasive technique phototrichogram, which allowed for individual hair shafts to be identified on a defined area of the scalp and for the growth phase of each shaft to be determined. Ten male subjects participated in the investigation and were observed at intervals of 4 to 6 weeks over the span of 8 to 14 years. Four of the subjects (labeled A, B, C, and D) exhibited normal hair growth, and the remaining six subjects (labeled E, F, G, H, J, K) exhibited characteristics of androgenetic alopecia [25].

## AA

We connect the Dobreva et at. (2018) model for AA with an estimate for the lengths of anagen and telogen deduced from information in the literature as such data from AA patients is not currently available. For the telogen phase of the HC we use as an estimate 3 months [26]. The anagen phase is understood to comprise six subphases, anagen I through anagen VI [1, 27], and in AA hair growth appears to stop some time before anagen V is reached [28]. Also, it is known that on the scalp the duration of anagen I through anagen V is relatively short, with the elongation of hair fibers occurring in anagen VI [1, 27] and that the duration of anagen phases I through V does not differ significantly between scalp follicles and follicles from other regions, such as under the temple [1, 29]. Thus, as an estimate for the length of anagen in AA, we use 22 weeks, which is the average length of anagen for follicles under the temple [29].

# Alignment of Data and Mathematical Model Simulations

Through manual parameter modulation and visual inspection of results, we identified that for the AGA case adjustment in the parameter $p_4$ (which is apoptosis of MKs) in the hair cycle model, can render good agreement to the data from the Courtois et al. (1995) study [25] for the lengths of anagen and telogen for the first and last cycle in both control subjects and subjects with AGA.

The AA case involved adjusting the parameters in the immune sub-system of the AA model with hair cycling, which capture the effects of IFN-$\gamma$, while the value for the parameter $p_4$ was set to the average of the $p_4$ values from the first and last cycles of the control subjects from the Courtois et al. (1995) study [25]. The alignment simulations were performed with source code developed in the programming and computational software MATLAB (MathWorks Inc., Natick, MA, USA) for a Runge-Kutta method of order five.

# Initial Screening for Parameter Importance

We use the Morris screening method [30] in order to perform initial screening of parameters with the goal to identify the model parameters of greatest importance in relation to the length of the anagen phase. These parameters are subsequently included in the uncertainty quantification analysis. The Morris screening method helps to determine the most important parameters (also referred to as inputs) by first computing for each parameter incremental ratios, called elementary effects. This is achieved by sampling each input at $k$ number of values called levels with a fixed distance between two consecutive levels of $\Delta$, and computing at different sample points

$$\mathbf{X}^i = (X_1^i, X_2^i, \ldots, X_j^i, \ldots, X_k^i), \quad i = 1, \ldots, N$$

one-step differences of the form

$$D_{X_j}^i = \left| \frac{Y(\mathbf{X}^i + \Delta \mathbf{e}_j) - Y(\mathbf{X}^i)}{\Delta} \right|.$$

In the above formula, $(\mathbf{X}^i + \Delta \mathbf{e}_j)$ equals the vector $\mathbf{X}^i$ except for its $j$th component that has been increased by $\Delta$, that is, it has been sampled at the successive level. Then, the incremental ratios, $D_{X_j}^i$, are averaged in order to compute the Morris sensitivity measure, $\mu_{X_j}$, which evaluates the overall influence of the input on the output [30, 31, 32, 33]. In this study, for the number of levels we used $k = 20$, and the distance between two consecutive levels is $1/(k-1)$.

## Uncertainty Quantification

Uncertainty quantification is a useful analysis tool in health-related problems, and it has been used, for example, in the study of pulmonary hemodynamics [34], epidemiological models for COVID-19 [35] and musculoskeletal models [36]. We perform uncertainty quantification with the Delayed Rejection Adaptive Metropolis (DRAM) method [37].

In the uncertainty quantification analysis, we included the parameters found to be most influential with the Morris screening method. For each parameter we generated a chain of size 10,000 iterations in order to obtain the posterior distribution, using a uniform prior distribution. For control and AGA subjects, we allowed for 50% variation around the nominal value of the parameters being analyzed. For the AA case, in order to keep the AA model in oscillatory regime for the uncertainty quantification analysis, we allowed for 50% variation in all analyzed parameters, except for $\alpha$, where we had 20% variation, and $k$, where we could only have 1% variation. The nominal parameter values are included in Table 1, with the exception of $p_4$, which is determined for each subject in the alignment of data and mathematical model simulations.

The DRAM algorithm employs Bayes' formula

$$\pi(X|Y) = \frac{\pi(Y|X)\pi(X)}{\pi(Y)},$$

which shows how the posterior parameter distribution $\pi(X|Y)$ is related with the prior parameter distribution $\pi(X)$ and the likelihood $\pi(Y|X)$ that data observations $Y$ are well approximated by the model given a set of parameter values $X$ [37, 31]. As a metropolis method, in DRAM a new candidate sample of parameter values is accepted only if it generates, compared to the current sample, higher likelihood that the data observations are well approximated by the model. However, there is a delay factored into the algorithm before rejection occurs, which in the case that the first proposed sample is not acceptable, allows for examining additional candidate samples. The adaptive feature of DRAM means that updates of the covariance for the posterior distribution are done using the history of the parameter chain being generated [37, 31].

To verify convergence for a DRAM chain, we use the Gelman-Rubin test. The first step in the test implementation is to generate $M$ number of chains for each parameter under consideration, where $M$ needs to be greater than or equal to 4. Then, from each chain we remove the number of iterations considered the burn-in period and divide the remaining number of values in each chain, $N$, in half in order to have $2M$ half-chains, that is chains of length $N/2$. The next step is to compute the potential scale reduction factor $\hat{R}$ for each parameter using $\hat{R} = \sqrt{\frac{v\hat{a}r(\theta)}{W}}$, where

$$W = \frac{1}{2M} \sum_{i=1}^{2M} s_i^2$$

and

$$v\hat{a}r(\theta) = \left(\frac{N-1}{N}\right) W + \frac{1}{N} B.$$

In the above expression for $W$, $s_i^2$ is the variance for each of the $2M$ chains. In the expression for $v\hat{a}r(\theta)$,

which is the marginal posterior variance for each parameter, $B$ is the variance between the $2M$ chains for that parameter. The value of $B$ is computed as follows

$$B = \frac{N}{M-1} \sum_{i=1}^{2M} (\bar{\theta}_i - \bar{\theta})^2,$$

where $\bar{\theta}_i$ is the mean within each of the $2M$ chains and $\bar{\theta}$ is the mean across all $2M$ chains [38].

Convergence of the DRAM chain for a parameter is implied by the corresponding potential scale reduction factor $\hat{R}$ having a value close to 1, and it is generally sufficient to use a threshold of 1.1. An $\hat{R}$ value near 1 indicates convergence because it means mixing and stationarity for the $M$ chains of length $N$. In regard to mixing, $\hat{R}$ value near 1 shows that the $2M$ half-chains have mixed, which implies that the $M$ chains of length $N$ have mixed as well. In regard to stationarity, $\hat{R}$ value near 1 shows that the first and second half of each of the $M$ chains traverses the same distribution [38]. The Gelman-Rubin test was performed with source code we developed in MATLAB.

The last part of the uncertainty quantification analysis we performed was computing Bayesian prediction intervals for the length of anagen based on 2000 values sampled from the posterior densities of the parameters under consideration. The prediction interval computation routines are part of the MATLAB DRAM codes discussed above.

## Global Sensitivity Analysis

We use the Sobol' method, which is based on analysis of variance (ANOVA) decomposition and helps to determine the total contribution of variation in a parameter to the output variance. The total variance $V$ of the output is distributed among all the effects through a sum of terms with increasing dimensionality, as follows

$$V(Y) = \sum_i V_i + \sum_{i<j} V_{ij} + \sum_{i<j<m} V_{ijm} + \cdots + V_{12\cdots k},$$

where

$$V_i = V[E(Y|X_i = x_i^*)], V_{ij} = V[E(Y|X_i = x_i^*, X_j = x_j^*)] - V[E(Y|X_i = x_i^*)] - V[E(Y|X_j = x_j^*)],$$

etc. The method of Sobol' estimates the total sensitivity index $S_{T_i}$ which combines all effects, first and higher order, involving parameter (or factor) $X_i$. A parameter is considered influential if it has a high total sensitivity index $S_{T_i}$ relative to other parameters [39].

The analysis procedure involves varying all parameters, simulating the model, collecting values for the outcome of interest, and calculating the total sensitivity index for each parameter. In our study, the outcome of interest is the length of the anagen phase. In addition, values for the parameters investigated with uncertainty quantification were sampled from the generated parameter chains. The remaining parameters were sampled from uniform distribution allowing for 30% variation around nominal value. As sample size, we used 25,000.

## Results

### Alignment of Data and Model Simulations

#### AGA

We identified that the parameter $p_4$, which captures the process of apoptosis of MKs, can render good alignment to the data for the lengths of anagen and telogen in both subjects with normal hair cycle and subjects with AGA. In normal conditions, in the first cycle $p_4 = 0.4994$ and $p_4 = 0.5086$ are the lowest and highest value, respectively, and in the last cycle $p_4 = 0.5079$ is the lowest value and $p_4 = 0.5269$ is the highest value. In AGA conditions, in the first cycle $p_4 = 0.5136$ and $p_4 = 0.5538$ are the lowest and highest value, respectively, and in the last cycle $p_4 = 0.5393$ is the lowest value and $p_4 = 0.5634$ is the highest value. In Figure 1, we present results for a subject with normal cycle, and in Figure 2, we show results for an AGA subject.

The simulations for some AGA subjects, where the value for $p_4$ is notably increased compared to the values in normal conditions, necessitated the initial condition for MK to be increased in order for cycling to be preserved. Specifically, we have $\xi(0) = 0.01$ for the first and last cycle of all subjects with normal hair growth (A, B, C, D) as well as the first and last cycle of 2 alopecia subjects (E and G) and only the first cycle of 3 alopecia subjects (H, J, K); we have $\xi(0) = 1.85$ for the last cycle of these 3 alopecia subjects. For alopecia subject F, the initial condition for $\xi$ is 1.85 for both the first and last cycle. The initial conditions for the remaining model components ($\eta_1$, $\eta_2$, $z_1$, $z_2$) are the same for all subjects in the first and the last cycle with values of $\eta_1(0) = 0.5$, $\eta_2(0) = 0.5$, $z_1(0) = 2$, $z_2(0) = 0.5$.

Based on these results from aligning the data and model simulations, the level of the parameter $p_4$ seems to outline cases of mild alopecia (both cycles of subjects E and G and the first cycles of subjects H, J, K) and severe alopecia (both cycles of subject F and the last cycles of subjects H, J, K). The highest $p_4$ level among the first and last cycles of subjects E and G and the first cycles of subjects H, J, K is $p_4 = 0.5405$. The lowest $p_4$ level among the first and last cycles of subject F and the last cycles of subjects H, J, K is $p_4 = 0.5579$. This suggests that with $p_4$ below 0.5405, presentation of AGA is mild, and as $p_4$ increases past 0.5405, severe presentation of AGA could be anticipated.

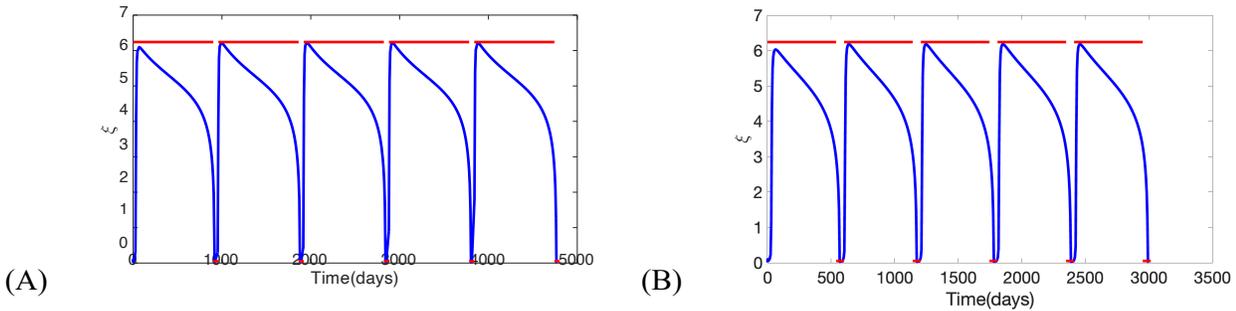

(A) (B)

Figure 1: Normal conditions. Subject A: blue - model, red - data. (A) First cycle, $p_4 = 0.4998$; (B) Last cycle, $p_4 = 0.5096$

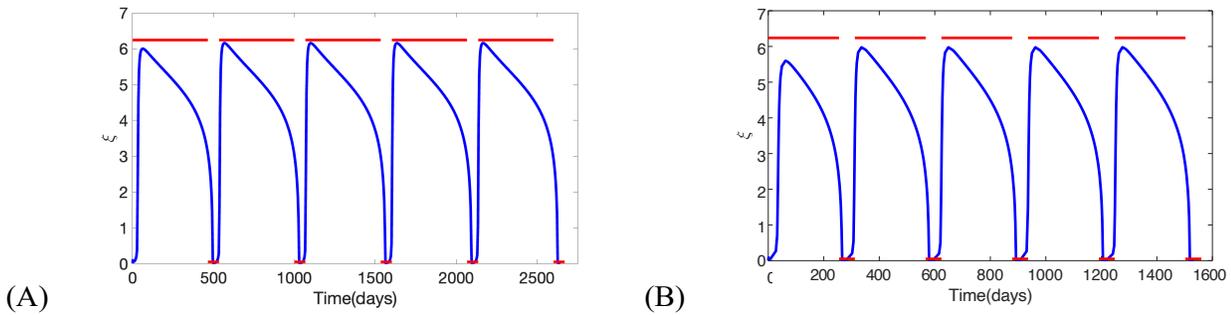

(A) (B)

Figure 2: AGA conditions. Subject E: blue - model, red - data. (A) First cycle, $p_4 = 0.5136$; (B) Last cycle, $p_4 = 0.5393$

## **AA**

The results for the AA case are shown in Figure 3. Alignment to the data for the lengths of anagen and telogen deduced from the literature was achieved with adjusting the parameters in the immune sub-system of the AA model with hair cycling, which capture the effects of IFN-$\gamma$. The value for the parameter $p_4$ was set to $p_4 = 0.5105$, which is the average of the $p_4$ values from the first and last cycles of the control subjects from the Courtois et al. (1995) study [25]. As the immune sub-system constitutes a reduction of a full model for the dynamics of immune system components driving AA, the parameters in the reduced model are combinations of parameters involved in the full immune model, as described in Dobreva et at. (2015) [20]. The parameters *a* and *b* in Equations (6) and (7) capture the effects of IFN-$\gamma$ as they contain the value *g*, which is the ratio of the production rate to degradation rate of IFN-$\gamma$ [20]. With setting *g* to $g = 6.5 \cdot 10^{-1}$, which in turn adjusts parameters *a* and *b* to $a = 1.5385 \cdot 10^{-4}$, $b = 8.1250$, respectively, we achieved good agreement to the data for the lengths of anagen and telogen deduced from the literature. In the immune sub-system, these values for the parameters *a* and *b* correspond to the state of disease characterized by elevated populations of the immune cells, which attack and kill MKs, causing hair loss.

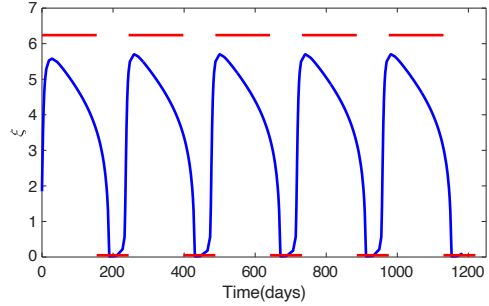

Figure 3: AA conditions. Blue - model, Red - lengths for anagen and telogen deduced from literature. $a = 1.5385 \cdot 10^{-4}$, $b = 8.1250$ ($g = 6.5 \cdot 10^{-1}$), $p_4 = 0.5105$

**Initial Screening for Parameter Importance**

The results from performing Morris screening for the first and last cycle for the control and AGA subjects as well as for the AA case are presented in Table 2. To generate these results, we ran the screening routine 10 times and took the average of the Morris sensitivity measure for each parameter over the 10 runs. As we can see, for all subjects with normal hair cycle (A, B, C, D) as well as the subjects with mild AGA in both cycles (E and G), the parameters *k*, $p_5$, $D_z$, $d_2$, $p_1$, $d_4$, $p_4$ are the most influential in relation to the length of the anagen phase in both the first cycle and the last cycle. In addition, for the three AGA subjects with mild AGA in the first cycle (H, J and K), these parameters are the most important for the duration of anagen in the first cycle, while in the last cycle, where severe AGA is exhibited, the parameters with greatest impact are $p_1$, $C_{prol}$, *k*, $p_4$. For the subject with severe AGA in both cycles (F), $p_1$, $C_{prol}$, *k*, $p_4$ are the parameters with greatest influence on how long anagen lasts in both the first and the last cycle. Also, for the AA case, the most influential in relation to the length of anagen are $p_1$, $C_{prol}$, *k*, $\alpha$, $p_3$ and $d_2$.

The parameter $p_1$ is the proliferation rate for MKs, and $p_3$ and $C_{prol}$ reflect inhibition of MK proliferation associated with regulatory molecules. The parameter $p_4$ reflects the process of apoptosis of MKs, and *k* and $p_5$ modulate the feedforward inhibition of MK apoptosis. Stem cell input to the MK population is represented by the parameter $\alpha$. The parameter $D_z$ is permeation constant of regulatory molecules. The parameters $d_2$ and $d_4$ reflect loss of signaling molecules ($\eta_2$) and regulatory molecules ($z_2$), respectively, from the DP compartment due to permeation into the MK compartment as well as leakage into the surrounding environment.

In light of the biological significance of the parameters identified as important, the findings from Morris screening indicate that MK proliferation and apoptosis, feedforward inhibition of MK apoptosis, permeation capacity of regulatory molecules and loss of signaling and regulatory molecules from the DP due permeation into the MK compartment as well as leakage into the surrounding environment are processes that strongly impact how long anagen lasts in normal conditions and mild AGA conditions. In addition, MK apoptosis and feedforward inhibition of MK apoptosis are crucial factors for the length of anagen in severe AGA conditions, and MK proliferation and permeation of signaling molecules from the DP into the MK compartment are important processes in severe AGA conditions as well as in the AA case. Another process important in severe AGA conditions and in the AA case, but not in the case of normal hair growth or mild AGA, is inhibition of MK proliferation associated with regulatory molecules.

Additionally, stem cell input to the MK population is a process important in the AA case, but not in normal and AGA conditions.

Table 1: Nominal parameter values. Note that: i) $p_4$ is not included since it varies among subjects and between the AGA and AA cases, ii) The immune model parameters apply only for the AA case.

| Hair cycle system | |
|---|---|
| Parameter | Value |
| $\alpha$ | 0.1 |
| $\beta$ | 0.01 |
| $p_1$ | 0.48 |
| $p_2$ | 0.1 |
| $p_3$ | 0.1 |
| $p_5$ | 0.32 |
| $C_{prol}$ | 1 |
| $c_1$ | 1 |
| $c_2$ | 1 |
| $k$ | 2.036 |
| $D_\eta$ | 0.5 |
| $D_z$ | 0.1 |
| $d_1$ | 2 |
| $d_2$ | 2 |
| $d_3$ | 2 |
| $d_4$ | 2 |
| Immune model | |
| $s$ | $3.4 \cdot 10^{-4}$ |
| $v$ | 0.3215 |
| $l$ | 0.6319 |
| $a$ | $1.5385 \cdot 10^{-4}$ |
| $b$ | 8.1250 |
| $w$ | 0.0837 |

Table 2: Morris screening results

| Normal conditions | | |
|---|---|---|
| Subject | First cycle | Last cycle |
| A | $k, p_5, D_z, d_2, p_1, d_4, p_4$ | $k, p_5, D_z, d_2, p_1, d_4, p_4$ |
| B | $k, p_5, D_z, d_2, p_1, d_4, p_4$ | $k, p_5, D_z, p_1, d_2, p_4, d_4$ |
| C | $k, p_5, D_z, p_1, d_2, d_4, p_4$ | $k, p_5, p_1, D_z, d_2, p_4, d_4$ |
| D | $k, p_5, D_z, d_2, p_1, p_4, d_4$ | $k, p_5, D_z, p_1, d_2, d_4, p_4$ |
| AGA conditions | | |
| E | $k, p_5, D_z, p_1, d_2, d_4, p_4$ | $k, p_5, p_1, D_z, d_2, p_4, d_4$ |
| F | $p_1, p_4, k, C_{prol}$ | $p_1, C_{prol}, k, p_4$ |
| G | $k, p_5, D_z, d_2, p_1, d_4, p_4$ | $k, p_5, p_1, D_z, d_2, p_4, d_4$ |
| H | $k, p_5, p_1, D_z, d_2, p_4, d_4$ | $C_{prol}, p_1, k, p_4$ |
| J | $k, p_5, p_1, D_z, d_2, p_4, d_4$ | $p_1, C_{prol}, p_4, k$ |
| K | $k, p_5, p_1, D_z, d_2, p_4, d_4$ | $C_{prol}, p_1, k, p_4$ |
| AA conditions | | |
| $p_1, C_{prol}, k, \alpha, p_3, d_2$ | | |

## Uncertainty Quantification

We conducted uncertainty quantification for the first and last cycle of control and AGA subjects using the DRAM method. We verified convergence for the DRAM chains with the Gelman-Rubin test where, $M = 10$ chains were generated for each parameter under consideration in order to compute the potential reduction factor $\hat{R}$. The parameters we included in the DRAM analysis are those highlighted as important with the Morris screening method. The Gelman-Rubin test results show $\hat{R}$ values near 1 (i.e., less than 1.1) for all parameter chains in all of the cases considered: control subjects' first and last cycle and AGA subjects' first and last cycle. This indicates that convergence was achieved for the chains of all parameters in all cases. The figures presenting the DRAM analysis results show densities plotted after averaging over the 10 DRAM chains generated for each parameter.

For each subject we compared between parameter densities in the first cycle and the last cycle. First, for control subjects, we found noticeable shifts in the peak of parameter densities in the last cycle compared to the first cycle - subject A: $p_4$, $p_5$, $k$, subjects B and C: $p_4$. For parameter $p_4$, the peak occurs at a higher value in the last cycle. For parameters $p_5$ and $k$, the peak occurs at a lower value in the last cycle. Also, for the control subject D for all parameters, a noticeable increase in the spread of parameter densities is observed in the last cycle compared to the first cycle. This larger spread points to more uncertainty in the parameter values and consequently the processes they represent.

Prominent shifts in the peak of the parameter densities in the last cycle compared to the first cycle are also present in almost all AGA subjects - subject E: $p_4$, subject G: $p_4$, $k$, subject H: $p_4$, subject J: $p_1$, $p_4$, $k$, subject K: $p_1$, $k$. For parameter $p_1$, the peak occurs at a lower value in the last cycle. For parameters $p_4$ and $k$, the peak occurs at a larger value in the last cycle. In addition, noticeable change in the spread of parameter densities in the last cycle compared to the first cycle is observed for subject E: $p_1$, $p_5$, $d_2$, $d_4$ (decrease of spread in last cycle compared to first) and subjects H, J and K: $k$ (increase of spread in last cycle compared to first). Notably, for the most serious AGA case (subject F), neither noticeable shifts in the peak nor changes in the spread of the distributions are exhibited.

Figure 4 presents the parameter densities for the AA case, and Figures 5 through 12, show comparison of parameter densities among the AA case and control and AGA subjects within the first cycle and the last cycle as well as comparison of all subjects' parameter densities between the first and the last cycle. We can see that the peaks of the $p_1$, $D_z$, $d_2$, $d_4$ densities for control and AGA subjects occur around a similar value, and this holds both in the first and the last cycle. In contrast, for the parameters $p_1$ and $d_2$, the densities for the AA case have a peak at a lower value than the control and AGA subjects. In addition, the spread of the distributions is similar, with in some instances only subject E having a wider distribution in the first cycle, and only subject D having a wider distribution in the last cycle.

For the parameter $p_4$, we can see that there is closer clustering of the densities of AGA subjects around a higher $p_4$ value in the last cycle compared to the first cycle. On the other hand, among the control subjects, the separation between the densities of subjects A and B and those of subjects C and D observed in the first cycle is still maintained in the last cycle. For parameter $p_5$ in the first cycle, while the density peaks occur at higher values for the control subjects A and B, the peaks for subjects C and D occur at values similar to those of AGA subjects. In the last cycle, for parameter $p_5$ the distinction of the densities of control subjects A and B is not as clear any more.

For parameter $k$, it can be more clearly seen in the first cycle that the peaks of the distributions for AGA subjects occur at lower values compared to control subjects, but this is no longer the case in the last cycle. In addition, for the AA case, the density for the parameter $k$ is much narrower than the densities for control and AGA subjects because $k$ needed to be varied only slightly in order to keep the model for AA in oscillatory regime for the uncertainty analysis. For the parameter $C_{prol}$, the peaks of the densities for the AGA subjects occur at a similar value in the last cycle. In comparison, for the AA case the peak of the density for $C_{prol}$ occurs at a higher value than for the AGA subjects.

We also computed Bayesian prediction intervals for the length of anagen based on 2000 realizations drawn from the posterior parameter densities, averaging over the 10 DRAM runs for each subject. The prediction intervals together with the data for anagen duration are presented in Table 3. The results show that in control subjects, in the last cycle compared to the first cycle, the lower and upper bounds of the prediction intervals are at values with higher relative change from the data measurement. This points to higher uncertainty in the length of anagen in the last cycle compared to the first cycle in control subjects.

Similar conclusion can be made for the AGA subjects where the upper prediction interval bounds are at values with higher relative change from the data measurement in the last cycle compared to the first, and for the lower prediction interval bounds, only in two of the six AGA subjects there is a slight decrease in the relative change from the data measurement in the last cycle compared to the first. Also, both in the first and the last cycle, there is greater uncertainty in the length of anagen for AGA subjects compared to control subjects indicated by lower and upper prediction interval bounds in AGA subjects at values with much higher relative changes from the data measurements compared to control subjects.

In addition, for the AA case, the lower prediction interval bound is at a value with high relative change from the literature estimate, while the higher prediction interval bound is at a value with very low relative change from the estimate. This only occurs for the AA case - in the subjects with AGA and control subjects, the lower prediction interval bound is at a value with smaller relative change from the data measurement than the upper prediction interval bound. This suggests that in AA, there is greater uncertainty associated with how much lower than the literature estimate (of 154 days) the length of anagen could be, while under AGA and normal conditions, there is greater uncertainty associated with how much higher than the data measurements the length of anagen could be.

## Global Sensitivity Analysis

Table 4 presents the results from conducting global sensitivity analysis with the Sobol' method. They show that for control subjects (A, B, C, D) as well as subjects with mild AGA (E and G in both cycles as well as H, J and K in the first cycle), the parameters of greatest importance are $d_1$, $d_3$, $c_1$, $C_{prol}$, $c_2$ and $D_\eta$. For subjects with severe AGA (F and H, J and K in the last cycle), the most important parameters are $d_4$, $d_2$, $d_3$, $d_1$, $\alpha$, $c_2$, $c_1$, $D_z$, $D_\eta$ and $k$, with the parameter $\beta$ also in the list for subjects J and K. For AA conditions, the most influential parameters are $p_4$, $d_4$, $d_3$, $d_1$, $c_2$, $c_1$, $D_z$ and $D_\eta$.

As explained earlier, the parameter $p_4$ captures the process of MK apoptosis, $C_{prol}$ reflects inhibition of MK proliferation associated with regulatory molecules, $\alpha$ represents stem cell input to the MK population, $k$ modulates the feedforward inhibition of MK apoptosis, $d_2$ and $d_4$ reflect loss of signaling and regulatory molecules, respectively, from the DP due to permeation into the MK compartment and leakage into the surrounding environment. The parameters $D_\eta$ and $D_z$ are the permeation constants of signaling molecules and regulatory molecules, respectively. The parameters $d_1$ and $d_3$ reflect loss of signaling molecules ($\eta_1$) and regulatory molecules ($z_1$), respectively from the MK compartment due to permeation into the DP compartment as well as leakage into the surrounding environment. The parameter $c_1$ represents synthesis of signaling molecules by MKs, and $c_2$ represents synthesis of regulatory molecules in the DP in response to the level of signaling molecules in the compartment. Finally, $\beta$ reflects loss of MKs due to factors other than apoptosis.

In light of the biological significance of parameters highlighted as influential by the global sensitivity analysis, the results indicate that under normal and mild AGA conditions, the processes with highest impact on the length of anagen are loss of signaling and regulatory molecules from the MK compartment due to permeation into the DP and leakage into the surroundings, synthesis of signaling molecules by MKs, synthesis of regulatory molecules in the DP, inhibition of MK proliferation by regulatory molecules, and permeation capacity of signaling molecules.

Under severe AGA conditions, all of these processes, except for inhibition of MK proliferation by regulatory molecules, are also important for the duration of anagen. In addition, under severe AGA conditions, the hair growth phase is impacted by feedforward inhibition of MK apoptosis, loss of signaling and regulatory molecules from the DP due to permeation into the MK compartment and leakage into the surroundings, permeation capacity of regulatory molecules, and stem cell input to the MK population. Loss of MKs due to factors other than apoptosis could also have a potential impact, as suggested by the parameter $\beta$ being classified as important in two out of the 4 patients with severe AGA.

Notably, the process of MK apoptosis emerges as influential for the length of the hair growth phase only in the AA case. Other processes that play an important role are permeation of regulatory molecules from the DP into the MK compartment (accompanied with some leakage into the surroundings), permeation of signaling and regulatory molecules from the MK compartment into the DP (accompanied with some leakage into the surroundings), production of signaling molecules by MKs, production of regulatory molecules in the DP, and permeation capacity of signaling and regulatory molecules.

# Discussion

In this study, we inform the mathematical model of Al-Nuaimi et al. (2012) for the human hair cycle with experimental data for the lengths of anagen and telogen available from the study of Courtois et al. (1995), which measured these lengths in male control subjects and subjects with androgenetic alopecia (AGA). We also connect our model for alopecia areata (AA) proposed in Dobreva et at. (2018) with estimates for the lengths of anagen and telogen in AA-affected hair follicles obtained through information in the literature. Subsequently, these data-informed models are used to perform parameter screening, uncertainty quantification and sensitivity analysis. We compare results within and between the control and AGA subject groups. We also compare the results from the AA case with the findings from control and AGA subjects.

The initial screening for important parameters with the Morris method shows that processes which strongly affect the length of anagen under all conditions (normal, mild AGA, severe AGA and AA) are MK proliferation and permeation of signaling molecules from the DP into the MK compartment. In terms of differences, we see that MK apoptosis, feedforward inhibition of MK apoptosis, loss of regulatory molecules from the DP due to permeation into the MK compartment, and permeation capacity of regulatory molecules are processes highly influential for the length of anagen in normal and mild AGA conditions, but not in severe AGA conditions and AA conditions. Also, inhibition of MK proliferation associated with regulatory molecules is impactful in severe AGA and in the AA case, but not in normal conditions and mild AGA conditions. Finally, stem cell input to the MK population is a process important in the AA case, but not in normal and AGA conditions.

The uncertainty quantification results for control and AGA subjects show that with respect to the duration of anagen, there is a lot more inter-subject variability in the parameters reflecting MK apoptosis and feedforward inhibition of MK apoptosis. On the other hand, we do not see as much inter-subject variability in the parameters reflecting MK proliferation, inhibition of MK proliferation associated with regulatory molecules, permeation capacity of regulatory molecules, and permeation of signaling molecules and regulatory molecules, from the DP into the MK compartment, accompanied with some leakage into the surroundings.

In addition, for control subjects and subjects with mild AGA in the first cycle or in both cycles, prominent differences were observed in the distributions of some parameters capturing processes that strongly influence the length of anagen between the first and the last cycle, such as MK apoptosis and feedforward inhibition of MK apoptosis. On the other hand, for the subject with most severe AGA (subject F), there did not seem to be any significant changes in the parameter distributions between the first and the last cycle. This is noteworthy since the two cycles are separated by a very large time span of at the least 8 years, and it suggests that the processes in the most severe AGA case seem to have a more established course, in the absence of external interventions. Future studies are needed to assess what would happens if treatment is applied.

The comparison of uncertainty quantification results for the AA case with the results from AGA and control subjects reveals that the density peaks of the parameters capturing MK proliferation and permeation of signaling molecules from the DP into the MK compartment occur at a lower value in the AA case than in control and AGA subjects. This indicates that in AA, proliferation of MKs could be expected to have lower level and the communication of the DP with MKs via signaling molecules to be weaker than in normal and AGA conditions. Also, the density peak of the parameter capturing inhibition of MK proliferation associated with regulatory molecules occurs at a higher value in the AA case than in the AGA subjects. This indicates that in AA, stronger inhibition of MK proliferation by regulatory molecules could be expected than in AGA.

The results for prediction intervals point to higher uncertainty in the length of anagen in the last cycle compared to the first cycle in control subjects. Additionally, both in the first and the last cycle, there is greater uncertainty in the length of anagen for AGA subjects compared to control subjects. Our findings also suggest that in the AA case, there is more uncertainty associated with how much the anagen duration could reach below the available estimate, while under AGA and normal conditions, there is more uncertainty in the possible anagen duration level above the data measurements. This suggests that in the absence of external intervention, for AA it is more certain that longer anagen could not be expected compared to control and AGA conditions.

The results from global sensitivity analysis with the Sobol' method, show that under severe AGA conditions, the processes with greatest impact on the duration of hair growth are loss of signaling and regulatory molecules from the MK compartment due to permeation into the DP, synthesis of signaling molecules by MKs, synthesis of regulatory molecules in the DP, permeation capacity of signaling molecules, feedforward inhibition of MK apoptosis, loss of signaling and regulatory molecules from the DP due to permeation into the MK compartment, permeation capacity of regulatory molecules, and stem cell input to the MK population. The findings regarding synthesis of regulatory molecules in the DP agree with the physiological understanding of how androgens affects hair follicles in AGA, namely that androgens cause alterations in the production of regulatory factors by the DP [40]. Some of these processes also play an important role in the AA case. However, the global sensitivity analysis classifies the process of MK apoptosis as highly influential for the length of anagen only in the AA case, but not for normal and AGA conditions.

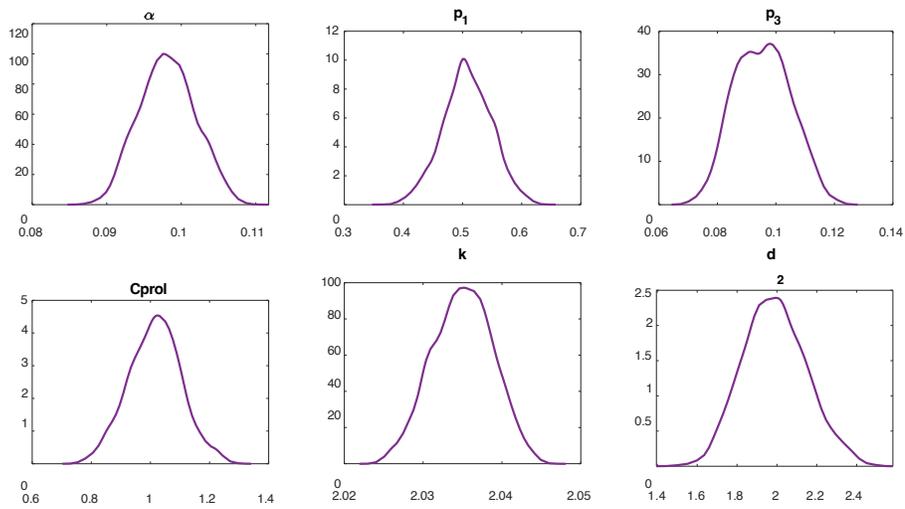

Figure 4: AA conditions.

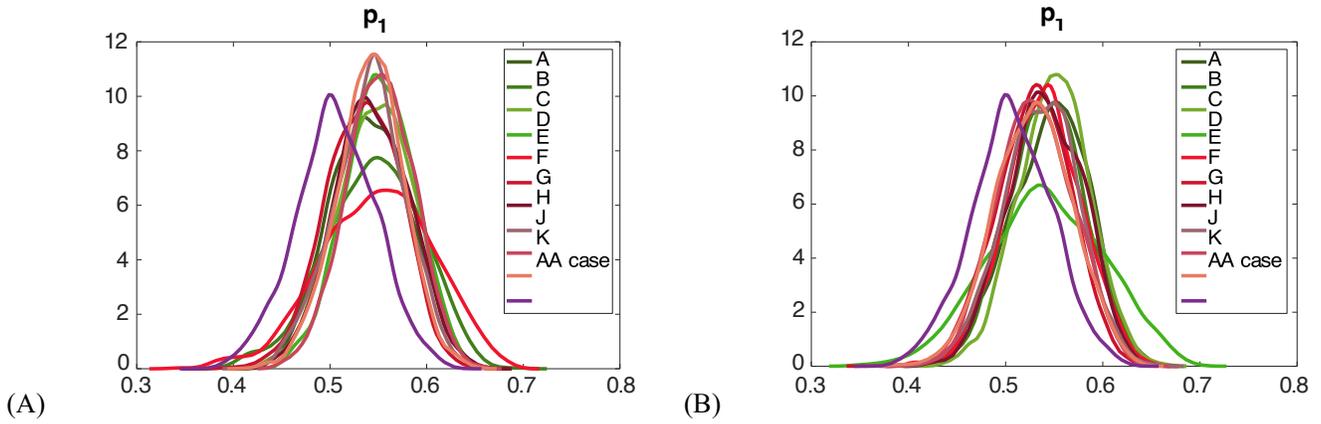

Figure 5: Parameter $p_1$ (A) First cycle (B) Last cycle

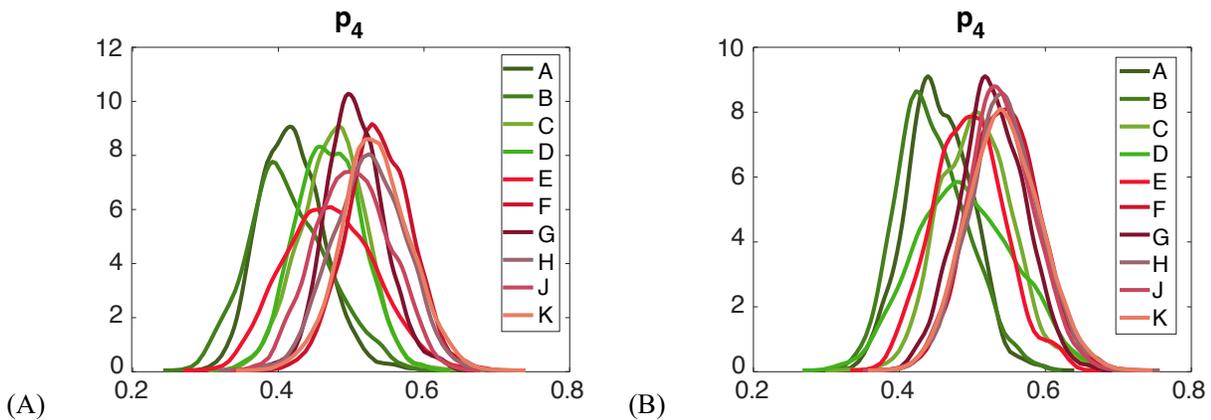

Figure 6: Parameter $p_4$ (A) First cycle (B) Last cycle

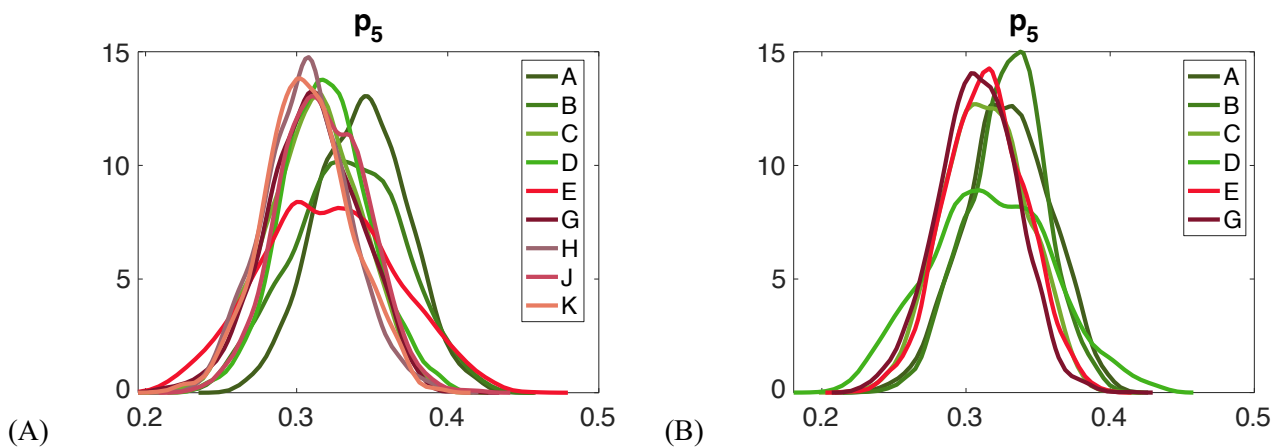

Figure 7: Parameter $p_5$ (A) First cycle (B) Last cycle

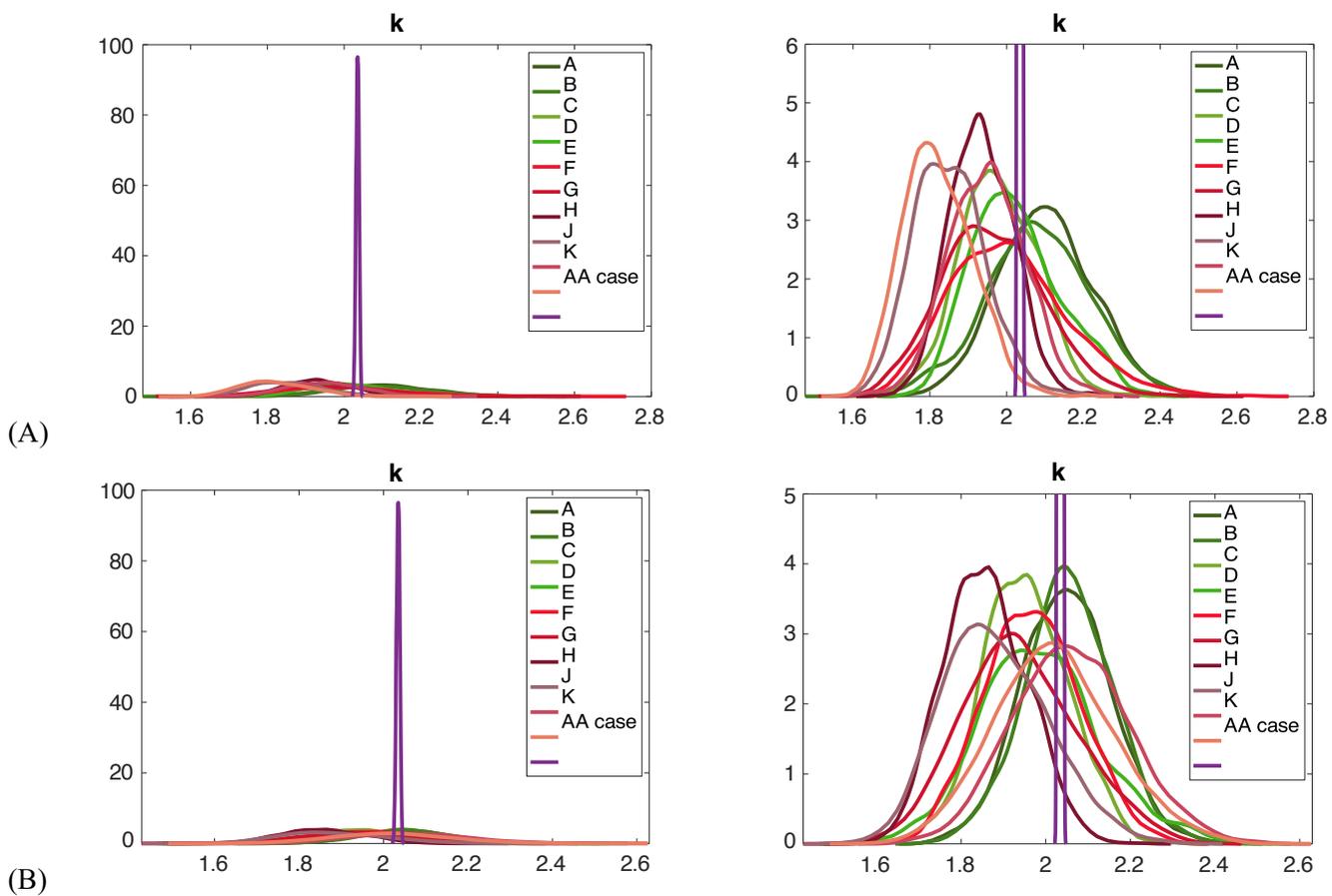

Figure 8: Parameter $k$ (A) First cycle. Closeup on right (B) Last cycle. Closeup on right

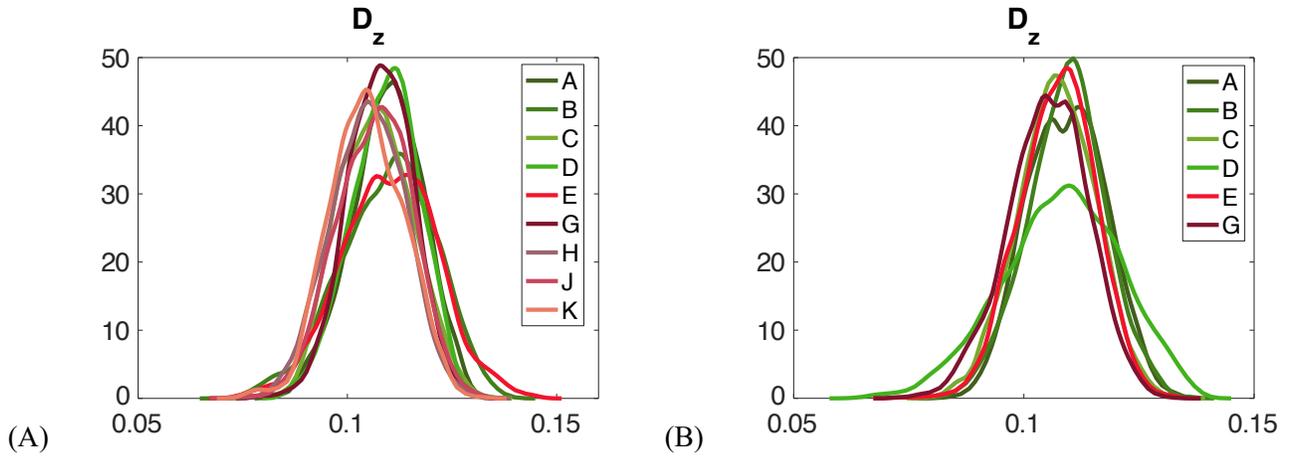

Figure 9: Parameter $D_z$ (A) First cycle (B) Last cycle

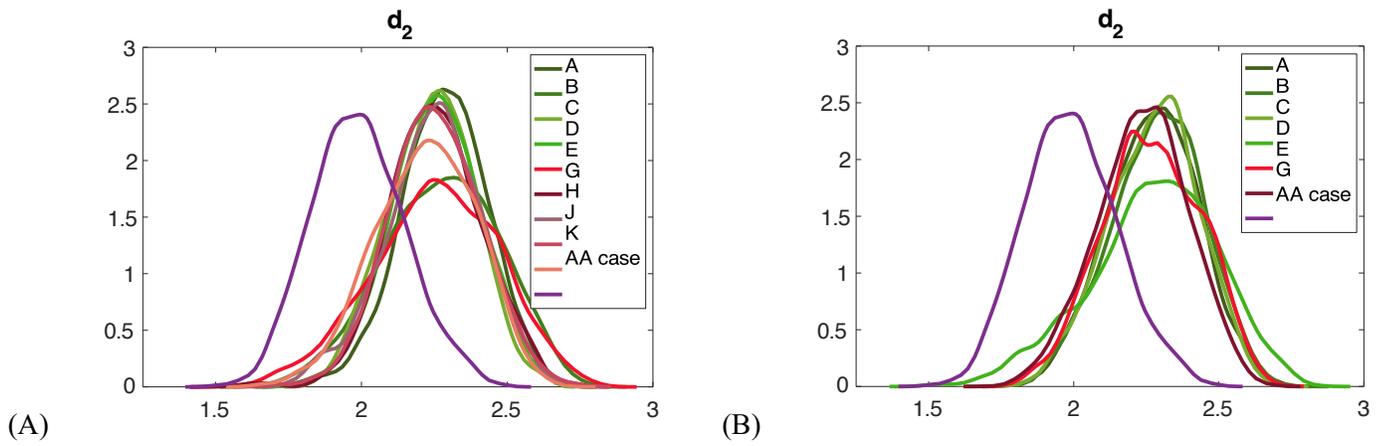

Figure 10: Parameter $d_2$ (A) First cycle (B) Last cycle

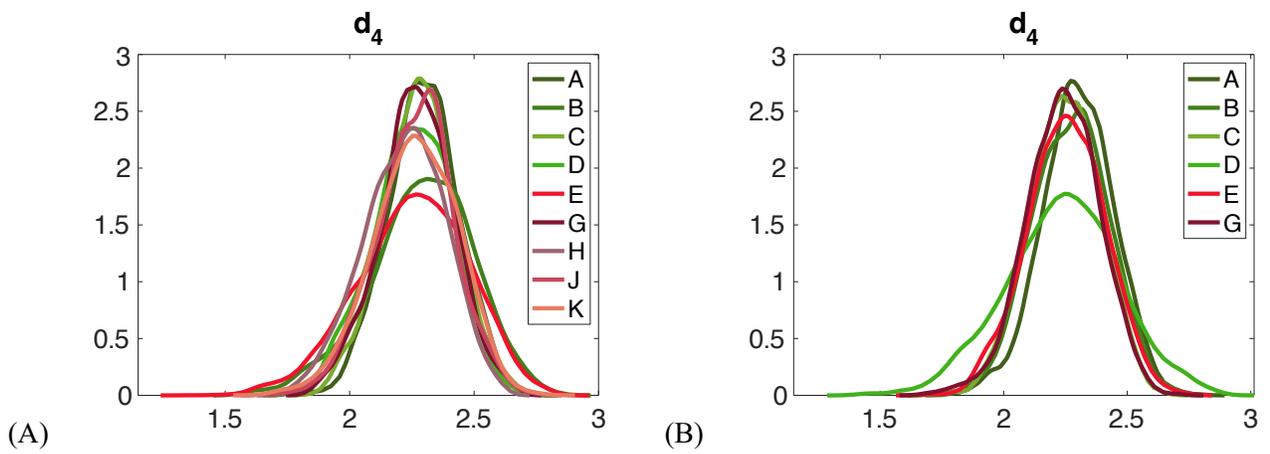

Figure 11: Parameter $d_4$ (A) First cycle (B) Last cycle

Table 3: Data and Bayesian prediction intervals for length of anagen in days

**Normal conditions**

| Subject | First cycle | | Last cycle | |
|---|---|---|---|---|
| | Data | Prediction interval | Data | Prediction interval |
| A | 903 | [783, 1105] | 540 | [394, 784] |
| B | 930 | [788, 1193] | 582 | [441, 834] |
| C | 552 | [377, 847] | 321 | [179, 553] |
| D | 573 | [408, 858] | 351 | [195, 615] |

**AGA conditions**

| Subject | First cycle | | Last cycle | |
|---|---|---|---|---|
| | Data | Prediction interval | Data | Prediction interval |
| E | 465 | [333, 718] | 255 | [103, 490] |
| F | 168 | [91, 385] | 129 | [58, 329] |
| G | 399 | [257, 674] | 228 | [84, 493] |
| H | 222 | [104, 484] | 126 | [68, 323] |
| J | 318 | [181, 556] | 168 | [87, 367] |
| K | 225 | [88, 496] | 150 | [67, 350] |

**AA conditions**

| Estimate from literature | Prediction interval |
|---|---|
| 154 | [0.31, 163] |

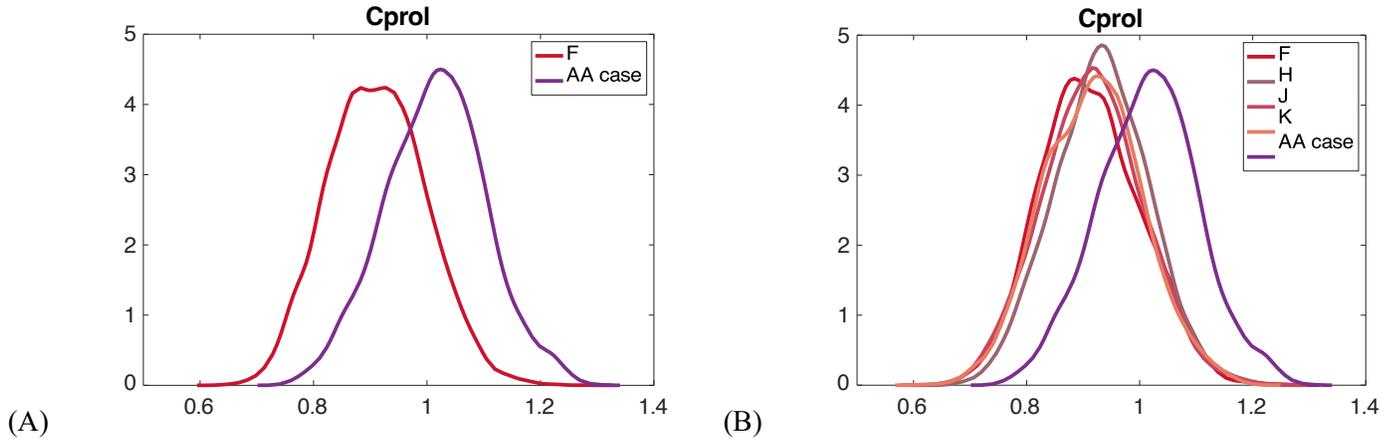

Figure 12: Parameter $C_{prol}$ (A) First cycle (B) Last cycle

Table 4: Sobol' sensitivity analysis results

| Normal conditions | | |
|---|---|---|
| Subject | First cycle | Last cycle |
| A | $d_1, d_3, c_1, C_{prol}, c_2, D_\eta$ | $d_1, d_3, c_1, C_{prol}, c_2, D_\eta$ |
| B | $d_1, d_3, c_1, C_{prol}, c_2, D_\eta$ | $d_1, d_3, c_1, C_{prol}, c_2, D_\eta$ |
| C | $d_1, d_3, c_1, C_{prol}, c_2, D_\eta$ | $d_1, d_3, c_1, C_{prol}, c_2, D_\eta$ |
| D | $d_1, d_3, c_1, c_2, C_{prol}, D_\eta$ | $d_1, d_3, c_1, c_2, C_{prol}, D_\eta$ |
| **AGA conditions** | | |
| E | $d_1, d_3, c_1, C_{prol}, c_2, D_\eta$ | $d_1, d_3, c_1, c_2, C_{prol}, D_\eta$ |
| F | $d_4, d_2, d_3, d_1, \alpha, c_2, c_1, D_z, D_\eta, k$ | $d_2, d_4, d_3, d_1, \alpha, c_2, c_1, D_z, k, D_\eta$ |
| G | $d_1, d_3, c_1, c_2, C_{prol}, D_\eta$ | $d_1, d_3, c_1, C_{prol}, c_2, D_\eta$ |
| H | $d_1, d_3, c_1, C_{prol}, c_2, D_\eta$ | $d_2, d_4, d_3, d_1, c_2, c_1, D_z, \alpha, D_\eta, k$ |
| J | $d_1, d_3, c_1, C_{prol}, c_2, D_\eta$ | $\alpha, d_2, d_4, d_3, d_1, k, c_2, c_1, D_z, \beta, D_\eta$ |
| K | $d_1, d_3, c_1, C_{prol}, c_2, D_\eta$ | $\alpha, d_2, d_4, d_3, d_1, k, c_2, c_1, D_z, D_\eta, \beta$ |
| **AA conditions** | | |
| | $p_4, d_4, d_3, d_1, c_2, c_1, D_z, D_\eta$ | |